\documentclass[12pt]{article}

\usepackage{amssymb}
\usepackage{graphicx}

\begin{document}
\title{A mean-field analysis of community structure in social and 
kin networks}
\author{Eric Durand$\dag$ \and Michael G.B. Blum$\ddag$ \and 
Olivier Fran\c cois$\dag$}
\maketitle

\vspace{0.5cm}

\noindent {\small $\dag$TIMB Department of Mathematical Biology, TIMC UMR 5525, Fac. 
M\'ed.,  Grenoble Universit\'es, F38706 La Tronche cedex, France}\par

\vspace{0.5cm}

\noindent {\small $\ddag$Department of Human Genetics, University of Michigan, 2017 
Palmer Commons, 100 Washtenaw Ave., Ann Arbor, MI 48109-2218, USA}\par

\vspace{0.5cm}

\noindent Corresponding author: {\tt olivier.francois@imag.fr}\par

\vspace{1cm}

\begin{abstract} 
We provide a mean-field analysis of community structure of social and 
biological networks assuming that actors are able to evaluate some tree-derived 
distance to the other actors and tend to aggregate with the less distant. We show that 
such networks have small components, and give exact descriptions for the probability 
distribution of a typical community size and the number of communities. In particular, 
we show that the probability distribution of the community size is well-approximated by a 
power-law distribution with exponent two. We illustrate the robustness of the mean-field 
analysis by comparing its predictions on previously studied social networks and 
biological data. 
\end{abstract}

\bigskip

\noindent {\bf Key-words: \quad} Community structure - Random trees - Coalescence - Distributional
recursions - power laws - kin networks - kin selection.

\clearpage

\section{Introduction}

Social networks have recently emerged as a paradigm of the complexity of human or 
animal interactions (Newman, 2003; Wasserman and Faust, 1994; Franck, 1998; Scott, 2000). 
Such networks are sets of actors with some pattern of 
contacts or interactions between pairs represented as edges in a graph.
It is widely assumed that most social networks show {\em community structure}, 
i.e., groups of strongly connected vertices, with few 
connections between groups (Girvan and Newman, 2002).
Community structure gives raise to a hierarchy of nested social relationships, 
which in turn can be thought of as a special kind of binary tree called a
{\it dendrogram} (e.g. Guimera et al., 2003; Arenas et al., 2004).

Algorithms that seek community structure in
graphs often attempt to reconstruct such a tree, and those that 
do so generally fall in two main categories: 
hierarchical clustering and edge removal (Scott, 2000; Girvan and Newman, 2002; Newman, 2004;
Radicchi et al., 2004).
In this study, we adopt a slightly different perspective on social networks, in which
the network itself derives from a hierarchical process represented as a tree.
The novelty is that the trees are considered as unobserved/hidden data, and 
there is no attempt to reconstruct them. Instead, the trees are viewed as random objects
which enable us to make predictions about the shape of the observed community structure. 

The network formation requires the actors to have an ability of assessing a (perhaps subjective) 
distance deduced from the tree (see Bogu$\tilde{\rm n}$a, 2004 and references therein for
similar postulates). Such distances are sometimes called {\it ultrametric}. Then 
the network evolves from the preferential attachment of each actor to the 
subset of her less distant actors. Here we present a mean-field analysis 
of community structure under this model. More 
specifically we describe the probability  distributions of a typical community size, 
and the number of communities in the network. 

In Section 2, we give a description of the mean-field theory for tree-derived 
networks, and show that the quantities of interest are involved in recursive distributional 
equations. In Section 3, we prove that the networks have small components, with community size 
depending logarithmically of the network size, while the number of communities depends linearly on the
network size. Then we study a variant of model with additional clustering, and obtain a number
of useful extensions of the previous results.  Section 4 illustrates and tests the robustness of the mean-field theory on two 
lists of examples, one from the social network literature, and the second from the sociobiology and 
ecology literature.

\section{Mean-field models}

\paragraph{Trees.} Ruling the basic principles of social network formation is an highly difficult
task. There is a large tradition in sociology for extracting community structure 
from a general network by cluster analysis (Scott, 2000; Newman, 2003). This method 
assumes a hierarchical organization of the network based on pair similarities 
(or distances).  Cluster analysis can generally be represented by a binary tree structure
(a dendrogram). Starting with $n$ vertices and  no edges, one adds one edge between 
the pair with the strongest similarities. Then, the two vertices are aggregated, and 
the distances to remaining vertices are recalculated. The process is 
iterated until all vertices aggregate. The connection between social network and trees 
is also exploited in reconstruction  algorithms that remove edges to the network
progressively (Girvan and Newman, 2003). 

In this study we assume that the network derives from a tree. The tree has internal branches 
that links the internal nodes to the root, 
and external branches that starts from the tips. The network is formed by going backward along 
external branches of the tree until a first ancestor is met. Edges are then 
drawn from each tip to all the descendants of the ancestor obtained in this process. 
We call this type of construction a {\it kin network} by analogy with biological networks 
where the tree represents a common genealogy. This construction is actually inspired from 
a biological process called {\em kin recognition} in which related individuals can recognize 
their kin, and attach preferentially to their closest relatives. This 
process forms the basis for the evolution of altruism (Hamilton, 1964). 
In the sequel this model is also referred to as the {\em perfect clustering} model
in contrast with an imperfect clustering model presented afterwards. 
 
Proposing models of random interactions between actors using learning and
rationality to evolve the structure is a standard approach in sociology and
sociobiology (see Skyrms and Pemantle, 2000 an references therein). Because it is more 
amenable to analysis, we consider a {\it mean-field} approximation of these interactions through the 
underlying tree process.  In the mean-field approximation, the tree is random. 
It starts with $n$ tips (the actors), adds one edge between a randomly chosen pair, and then coalesces the two 
tips into an ancestor. This model is often called a coalescent tree (see Aldous, 1999),
and arises as a robust approximation of the neutral genealogical process in population 
genetics (Kingman, 1982). In analogy with studies of genetic polymorphism, we never attempt to reconstruct 
the tree. The coalescent model is used as a basis for analyzing data such as community sizes or 
number of communities in a network. Still in analogy with population genetics, the coalescent tree
 may also serve as a model for testing the null-hypothesis that social networks evolve under random/neutral 
interactions.

\paragraph{Recursive definition of random trees.} Random coalescent trees share the same 
topology as other well-studied branching processes (Yule, 1924; Harding, 1971; Aldous, 2001).
Considering $n$ tips, these trees have the particular property that the size $L_n$ of the 
left sister clade at the basal split of the tree has uniform distribution over the set $\{ 
1, \dots, n-1 \}$ $$
{\rm P} ( L_n = \ell ) = \frac1{n-1} , \quad \ell = 1, \dots, n-1,
$$
and this property is also valid within each subtree. From this, Aldous (1996) proposed 
a  recursive definition of dendrograms through a {\it split distribution}, 
the distribution of the left sister clade given the 
size of the parent clade. The connection been random trees and recursive structures 
have been exploited by Blum and Fran\c{c}ois (2005a) to prove results about minimal clades
in the neutral coalescent. Their results can be rephrased to say that the out\-degree 
${\rm out}_n$ of an arbitrary vertex in a network with perfect clustering
has a power-law distribution with exponent $\alpha = 3$. More precisely we have
$$
{\rm P} [ {\rm out}_n = x ] = \frac{4}{x(x+1)(x+2)}, \quad x = 1, \dots, n-2,
$$
and ${\rm P} [ {\rm out}_n = n - 1 ] = 2/n(n-1)$, where $n$ is the network size.
Power-laws are not surprising in this context since this parallels similar 
results for (perhaps undirected) networks with incremental construction such as 
the Albert-Barabasi model (Albert and Barabasi, 2002) or the Price model (Price, 1965). 
See also (Newman, 2003) and (Durrett, 2006).

\paragraph{Imperfect clustering.} Communities in real social networks may sometimes consist 
of two or several subcommunities whereas this property is partly missing in 
the perfect clustering model for which each community is a clade subtended by and edge that 
connects a tip. Therefore we consider a modification of the basic model that tolerates {\em 
imperfect clustering} without modifying the underlying tree model. In the imperfect 
clustering model, communities may sometimes arise from the random coalescence of two 
previously formed clusters in addition to those created from the perfect clustering 
process. We also assume that the random clustering events occur during the construction
process at a rate $p$, called the {\em clustering rate} (See Fig.1).

\paragraph{Distributional recursions.} 
In this work the recursive definition of random trees is exploited to study the 
mathematical properties of community structure under the mean-field model. 
This is done by using distributional recursions. 
We call a {\it typical community} the network cluster that contains the leftmost tip 
in the underlying tree (the tip labelled 1). Because in the 
mean-field model the $n$ actors play exchangeable roles, studying the leftmost actor's 
community amounts to study an arbitrary community. We denote the community size by $S_n$ 
for $n$ the total network size. Obviously, we have $S_2 = 2$, and  
$S_3 = 3$. To give a recursive definition of $S_n$ (and then forget the tree), 
let us split the tree at the root so that two sister clades of sizes $L_n$ and 
$R_n = n - L_n$ are obtained, and let $I_n = \min(L_n, R_n)$. The community size $S_n$ 
can be recursively defined by $S_n = n$ if $I_n = 1$, otherwise $S_n = S_{L_n }$. 
In this definition, the replicates of $L_n$ are recursively sampled from 
the uniform distribution. The above set of recursive 
equations basically translates the idea of self-similarity and the scale-free 
property for a typical community size, but it also provides us with an efficient 
simulation algorithm for the probability distribution of $S_n$ that avoids the simulation 
of the tree itself. Sets of recursive distributional equations 
such as those described here also appear in computer science and are natural in the 
analysis of random  divide-and-conquer algorithms (R\"osler, 2001; 
Hwang and Neinninger, 2002; Blum and Fran\c{c}ois, 2005b). 
Regarding the number of communities $N_n$ we have $N_2 = N_3 = 1$. 
Like $S_n$, $N_n$ is involved in a set of recursive distributional equations. 
The number of communities can actually be defined as  
$N_n = 1$ if $I_n = 1$, and otherwise $
N_n =
N_{L_n } + N^*_{R_n}
$
where $N^*_n$ denotes an independent copy of $N_n$. Again the above recursive equations 
provide an efficient simulation algorithm for the probability distribution of $N_n$.

Turning to the model with imperfect clustering, the equations 
for the community size $S_n$ change as follows. We now have 
$S_n = n$ if  
$I_n = 1$,
otherwise
$$
S_n = \left\{ \begin{array}{ll}
n & {\rm with~probability~}  p \\    
S_{L_n } & {\rm with~probability~}  q    \\
\end{array}\right.
$$
where $q = 1-p$. Regarding the number of communities we have
$N_n = 1$ if $I_n = 1$,
and otherwise, 
$$
N_n = \left\{ \begin{array}{ll}
1 & {\rm with~probability~}  p \\    
N_{L_n } + N^*_{R_n} & {\rm with~probability~}  q     \\
\end{array}\right.
$$
Community structure in the mean-field model and recursive computations for $N_n$ are 
illustrated in Fig.1 where an example with $n = 12$ vertices 
and perfect and imperfect clustering is presented.

\begin{figure}
\label{fig:0}
\begin{center}
\includegraphics[width = 12cm]{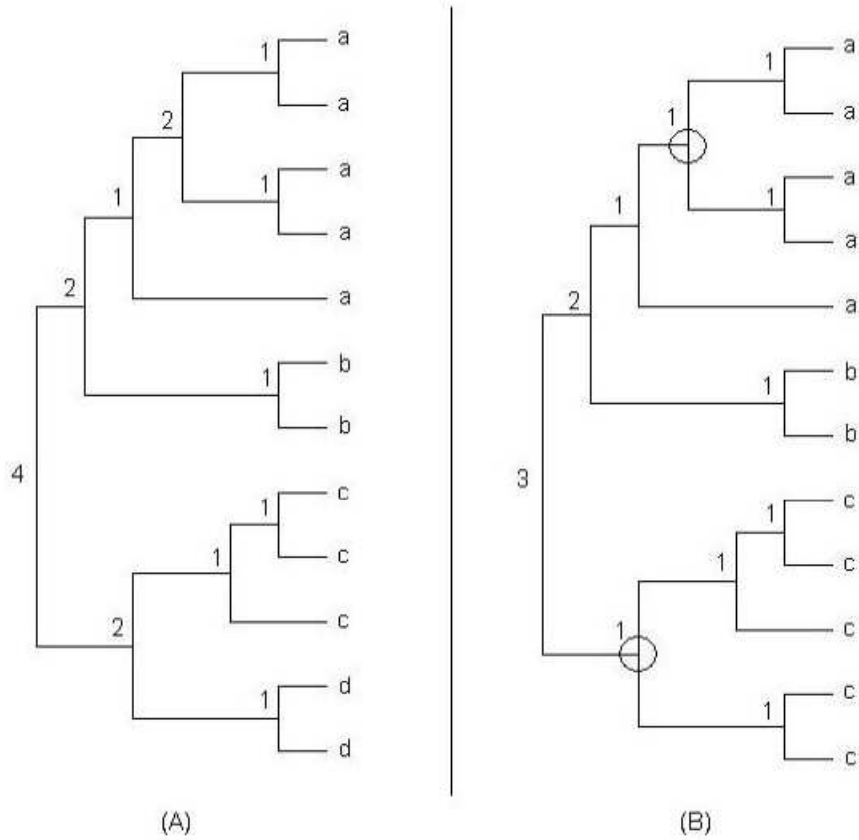}
\caption{\em Recursive computations of $N_n$ for $n = 12$. (A) Perfect clustering: the network 
has 4 communities. (B) Imperfect clustering: Two clustering events occur and are symbolized by 
circles. The network has 3 communities. Letters at the tips stand for community labels.}
\end{center}
\end{figure}

\section{Community size and the number of communities}

\subsection{Typical community size} 

\paragraph{Probability distribution.} 
We consider the probability distribution of $S_n$, and denote it by 
$p_n (x)$ $= {\rm P}(S_n = x)$ for all $2 \leq x < n$.
Then, for large $n$, we have $$p_n(2) \sim \frac{e - 2}e,$$ and for all $x \geq 
3$, we have
\begin{equation}
 p_n (x) \sim (-1)^{x+1}  2 x! \, \frac{ e^{-1} - e(x)}{x - 1}, \quad {\rm as~} n \to \infty
\label{eq:ps}
\end{equation}
where $ e(x)= \sum_{k=2}^x (-1)^k/k!$ is defined as the exponential sum function 
$e_x(z)$ at the point $z = - 1$. In particular, for $x = 3$, we obtain that 
$p_n(3) \approx 2(3-e)/e$, etc.

The key argument for obtaining the above probability distribution 
is the use of the recursive equations defining $S_n$. From the 
formula of conditional probabilities, we see that the $p_n(x)$'s are involved in 
sets of recursions of the following form 
$$
p_{n+1}(x) = (1 - \frac1n) p_n(x) + \frac1n p_{n-1}(x) 
$$
for $n \geq x+1$. For $x = 2$, the initial values are $p_2(2) = 1$ and $p_3(2) = 0$. 
For $x \geq 3$, the recursions start from $p_x(x) = 2/(x - 1)$ and $p_{x+1}(x) = 0$. 
The proof of the result stated in Eq. \ref{eq:ps} follows from considerations about  
the differences $u_{n+1} =  p_{n+1} - p_n$ and elementary calculations. Numerical 
computations show that the approximations given by Eq. \ref{eq:ps} are accurate for 
$n \geq 20$.  For large $x$ and $n$, there is a perfect 
agreement with a power-law distribution of exponent $\alpha = 2$ 
$$
p(x) \sim \frac{2}{(x+1)(x-1)}, \quad {\rm as~} x \to \infty \, .
$$
Numerical computations (not reported) show that the large $n$ - large $x$ approximations 
are accurate for $x \geq 25$ and $n \geq 100$.

\bigskip

\paragraph{Expected community size.} The above result suggests that under the 
mean-field model with perfect clustering, the size of a community grows to infinity as the 
number of actors increases. Here we give a more precise result, which states 
that the growth is in fact very slow. Considering the expected size, we denote 
$s_n = E[S_n]$, and obtain that  
$$
s_n \sim 2 \log(n) , \quad {\rm as~} n \to \infty. 
$$
In other words the kin networks studied here have small components. The sets of recursions
for $s_n$ are similar to that obtained for $p_n$. Actually we have
$$
s_{n+1} = (1 - \frac1n) s_n + \frac{s_{n-1}}n  + \frac{2}n, \quad n \geq 3,
$$
and the initial values are $s_2 = 2$ and $s_3 = 3$. The equation involving 
the difference $u_n$ can also be solved, and leads to $u_n = 2 A_{n-1}/n!$ where 
$A_{n-1}$ is the alternating factorial sum (Sloane's sequence number A005165 in 
EIS)). For large $n$, we obtain that 
$
u_n \sim 2/n
$
using (Abramowitz and Stegun 1970) and this leads us to the conclusion 
that $s_n \sim 2 (\gamma + \log(n))$ with $\gamma$ the Euler constant. To 
establish comparisons with the sequence $2 \log n$,
we find that $s_{n}/\log n \approx 2.04$ for $n = 1000$, 
and $s_{n}/\log n \approx 2.02$ for $n = 100,000$.

More generally, if we let $s_n^k = E[S_n^k]$, $k\geq 1$, denote the $k$th moment 
of the community size $S_n$, then for large $n$ and $k\geq 2$, we obtain that 
$$
s_n^k \sim \frac{2k}{k-1}n^{k-1} , \quad {\rm as~} n  \to \infty .
$$
In particular we have  $s_n^2 \sim 4 n$, and the variance of $S_n$ grows as 
4$n$.  

\subsection{Number of communities}

\paragraph{Probability distribution of the number of communities.} The probability distribution of 
$N_n$ can be computed exactly by solving a triangular system. If we let $\pi_n(x) =$ P$(N_n = x)$ for all integer $x$, we have
$\pi_n(1) = 2/(n-1)$, and 
$$
\pi_n(x) = \frac{1}{n-1} \sum_{\ell = 2}^{n - 2} 
\, \sum_{y=1}^{x-1} \pi_{\ell} (y) \pi_{n - \ell} (x - y)
 \quad 1 \leq x \leq \lfloor n/2 \rfloor \, . 
$$

\paragraph{Expected Number of communities.}  The expected number of communities is proportional 
to the number of vertices in the network 
$$
e_n = E[N_n] \sim  cn  ,\quad c =\frac{1-e^{-2}}{4} = 0.216\dots 
$$
for large $n$. From the recursive definition and a basic use of conditional 
probabilities, we obtain subsets of recursive equations for all the moments of $N_n$. In 
particular, the expected number of communities satisfies the following
recursion
$$
e_{n+1} = \left( 1 - \frac1n \right) e_{n} + \frac2n e_{n-1}, \quad  n \geq 3
$$
where the inital values are $e_2=e_3=1$. In the appendix, we show that this leads to
$$
e_n= (1-e^{-2}) (n + 2)/4 + O(2^n/(n-1)!) \, .
$$

\paragraph{Convergence in probability.} To see a convergence in probability 
result, remark that $t_n = E[N_n^2]$ solves the recursion
$$
t_{n+1} - t_n =  - ( t_n - t_{n-1} )/n + t_{n-1}/n +  2 ( r_{n+1} - r_{n} )/n ,
$$
with the residual term $r_n$ equal to 
$
r_n = \sum_{i=2}^{n-2} e_i e_{n - i}. 
$
Having proved $e_n \sim c n $ for large $n$, we can check that the residual 
difference term is equivalent to
$
 r_{n+1} - r_{n}  =  c^2 n^2/2 + o(n^2)
$. 
This leads to 
$ E[N_n^2] = c^2 n^2 + o(n^2)$
which can be translated into a convergence in probability result ($N_n/n \to c$)
by a standard  application of the Chebischev's inequality.

\subsection{Imperfect clustering}

Assuming imperfect clustering at rate $p$ modifies the recursions for $S_n$ and 
$N_n$, and complicates their mathematical analysis. Nevertheless, the main results 
can be summarized as follows (see the Appendix for details).

\paragraph{Community size.}  Assume that the clustering rate is positive $p>0$.
For $1<x \leq n$, we let $p_n(x) = $ P$(S_n = x)$. Then, we have 
for all $n \geq x+1$
\begin{equation}
p_{n+1}(x) = (1 - \frac1n) p_n(x) + \frac{q}{n}  p_{n-1}
\label{eq:snp}
\end{equation}
where $p_x(x) = p + 2q/(x-1)$ and $p_{x+1}(x) = 0$. 

To discuss the probability distribution of $S_n$ under imperfect clustering, let 
us distinguish the case $x=2$ from the general case.
For $x = 2$, Eq. \ref{eq:snp} starts with the initial values $p_{2}(2)=1$ and $p_{3}(2)=0$.
We set
$$
 I(p) =  q \int_{0}^{1}   y^2 e^{qy} (1-y)^{-p} dy \, .
$$
When $n$ grows to infinity, we obtain that
$$
p_{n}(2) \sim \frac{e^{-q}}{\Gamma(q)} I(p) n^{-p} , \quad n \to \infty
$$
For $p \to 0+$, we have
$$
\frac{e^{-q}}{\Gamma(q)} I(p) = \frac{e-2}{e} + a p + O(p^2).
$$
With $a = e(\gamma-1 +e^{-1}{\rm E}_{\rm i}(1,1)) = 0.0639$ (${\rm E}_{\rm i}$
is the exponential integral, see Abramowitz and Stegun, 1970). For $x>2$, we denote 
$f_{n} = p_{n+x}(x)$, and we have 
\begin{equation}
\label{eq:fx} (n+x-1)f_{n} = (n+x-2)f_{n-1} + q f_{n-2},
\end{equation}
where $f_{0} = p + 2q/(x-1)$ and $f_{1}=0$.
Using notations similar as above, we set 
$$
I_{x}(p) = q(2+(x-3)p) \, \int_{0}^{1} y^x e^{qy}( 1-y)^{-p}dy, 
$$
and, we obtain that for large $n$, 
$$
p_{n}(x) \sim \frac{e^{q}}{(x-1)\Gamma(q)} I_{x}(p) n^{-p} \, .
$$
The expected value $s_n = E[S_n]$ solves the following  recursion  
$$
s_{n+1} = \left( 1 - \frac1n \right) s_n + \frac{q}n  s_{n-1}  + 2 \left( p + \frac{q}n \right) 
$$
where the initial values are $s_2 = 2$ and $s_3 = 3$. The solution satisfies 
$$
s_n \sim \frac{2p}{1 + p} n \, , \quad n \to \infty
$$ 
for $0< p \leq 1$.


\paragraph{Number of communities.}  Regarding the number of communities, 
we let $\pi_n(x) =$ P$(N_n = x)$. Then the distribution $\pi_n$ can be calculated 
using triangular induction as follows 
$$
\pi_n(1) = p  + \frac{2q}{n-1}, 
$$ 
and 
$$
\pi_n(x) = \frac{q}{n-1} \sum_{\ell = 2}^{n - 2} 
\, \sum_{y=1}^{x-1} \pi_{\ell} (y) \pi_{n - \ell} (x - y) \, , 
 \quad 2 \leq x \leq \lfloor n/2 \rfloor \, . 
$$
We can use the above set of recursive equations to compute exact distributions up to network 
sizes greater than $n = 500$. In addition these equations enable us to obtain maximum likelihood estimates 
of the clustering rate $\hat p$ using either basic grid or more elaborate 
dichotomic searches. 
 
\begin{figure}
\begin{center}
\label{fig:zach}
\includegraphics[width = 6cm]{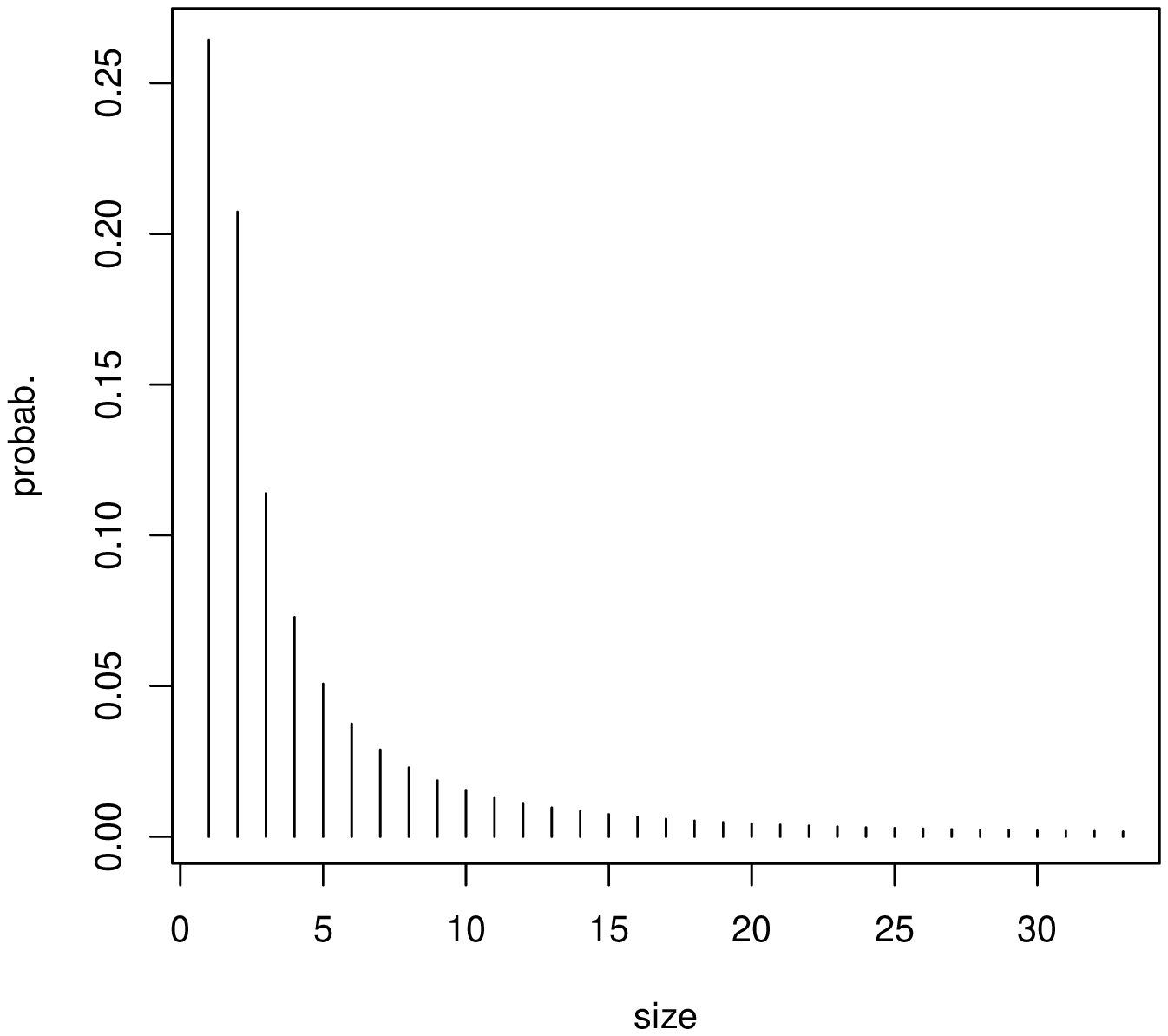}
\includegraphics[width = 6cm]{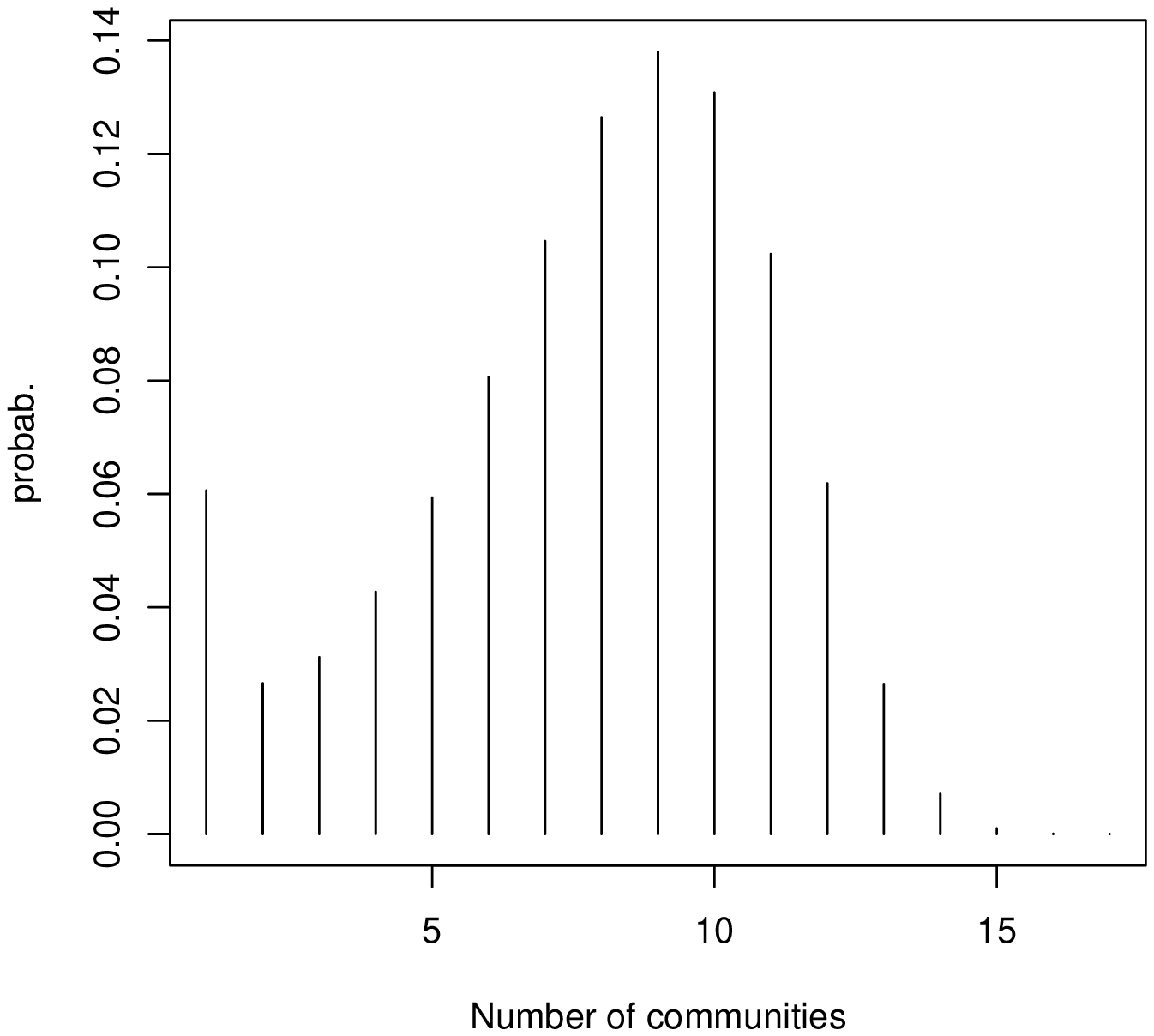}
\includegraphics[width = 6cm]{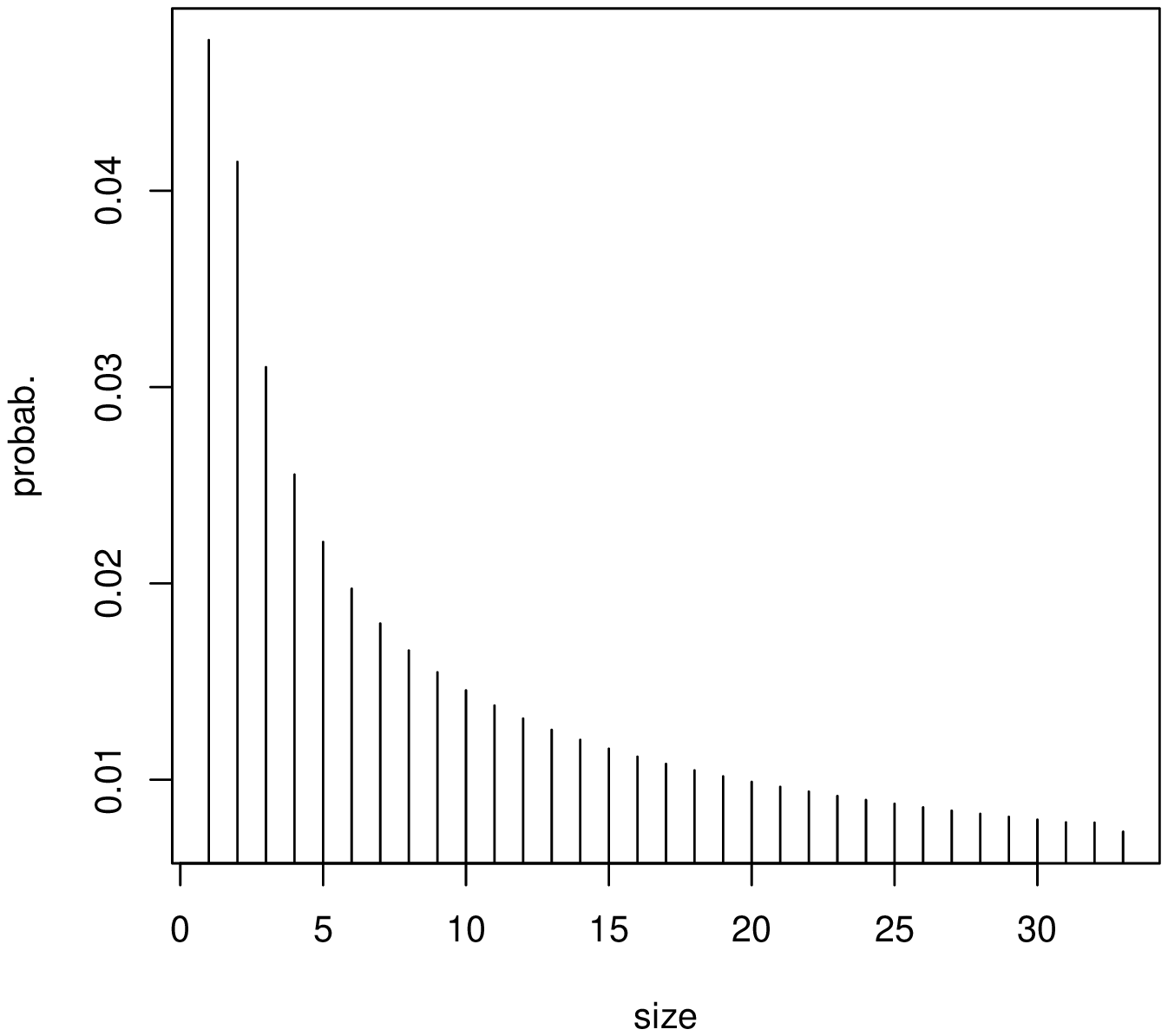}
\includegraphics[width = 6cm]{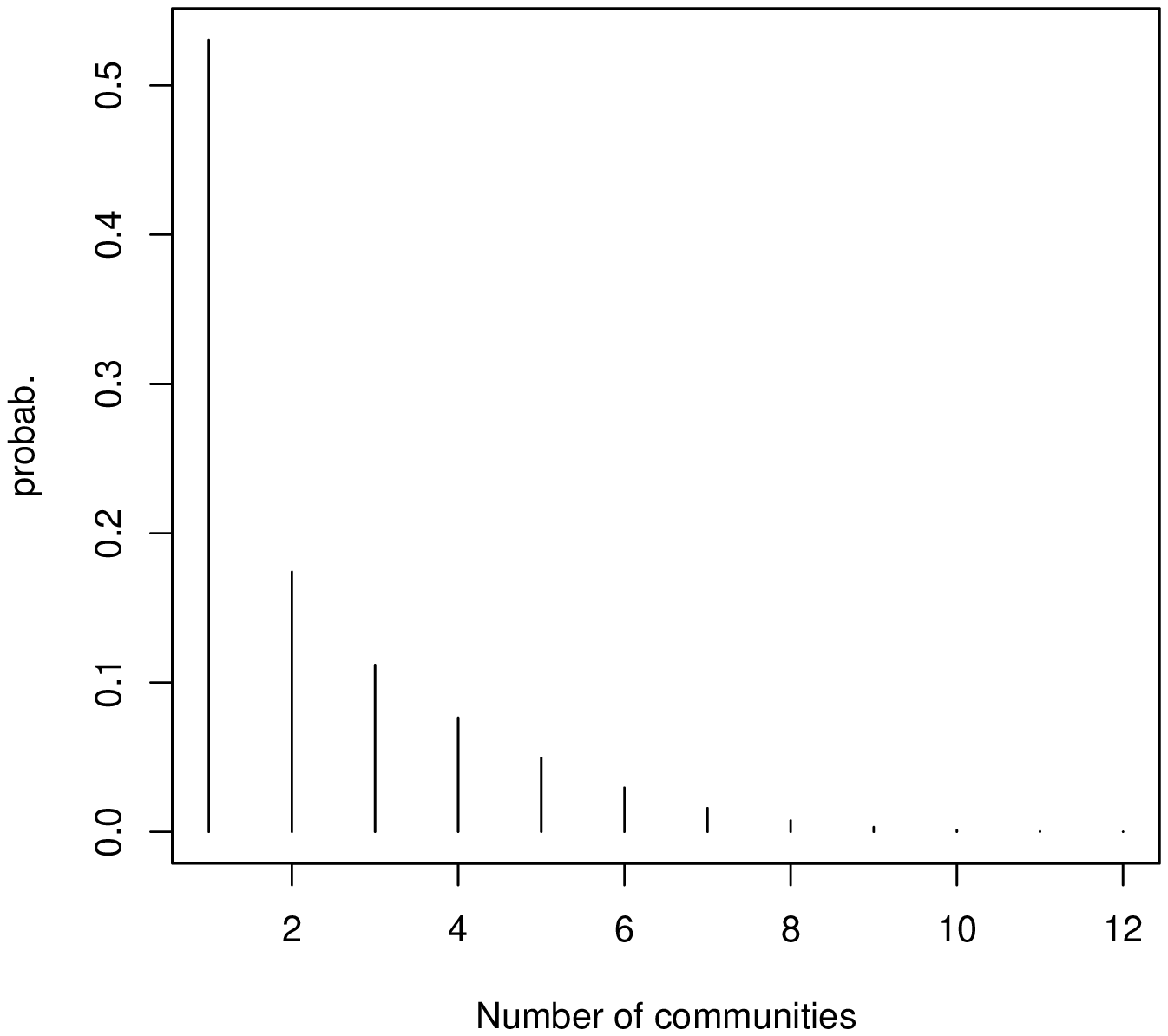}
\caption{Probability distribution of $S_{34}$ (left) and $N_{34}$ (right) 
for $p = 0$ (up) and $p = 0.5$ (down)  corresponding to Zachary's friendship 
network. Data from (Zachary 1977).}
\end{center}
\end{figure}

\begin{figure}
\begin{center}
\label{fig:foot}
\includegraphics[width = 6cm]{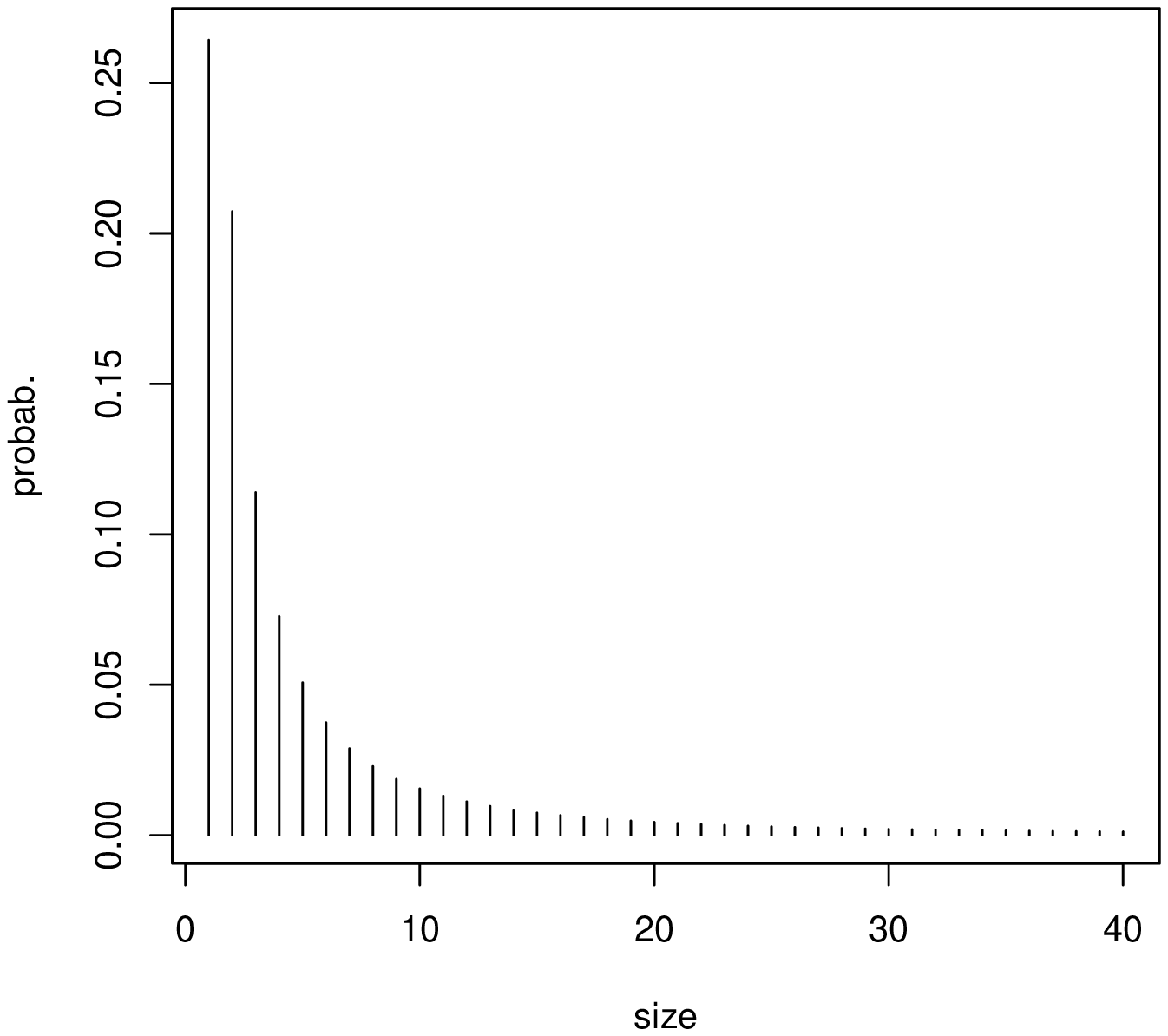}
\includegraphics[width = 6cm]{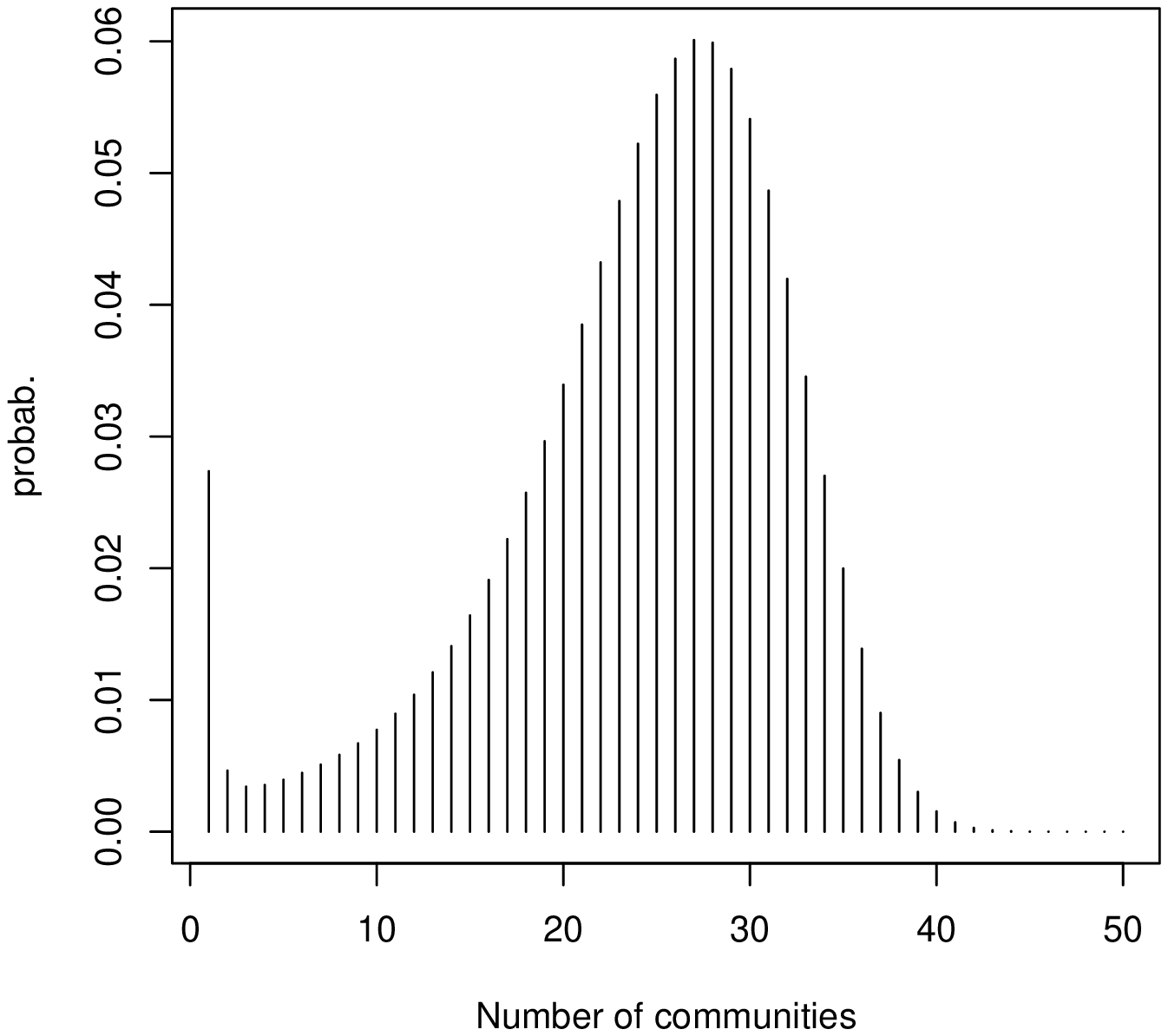}
\includegraphics[width = 6cm]{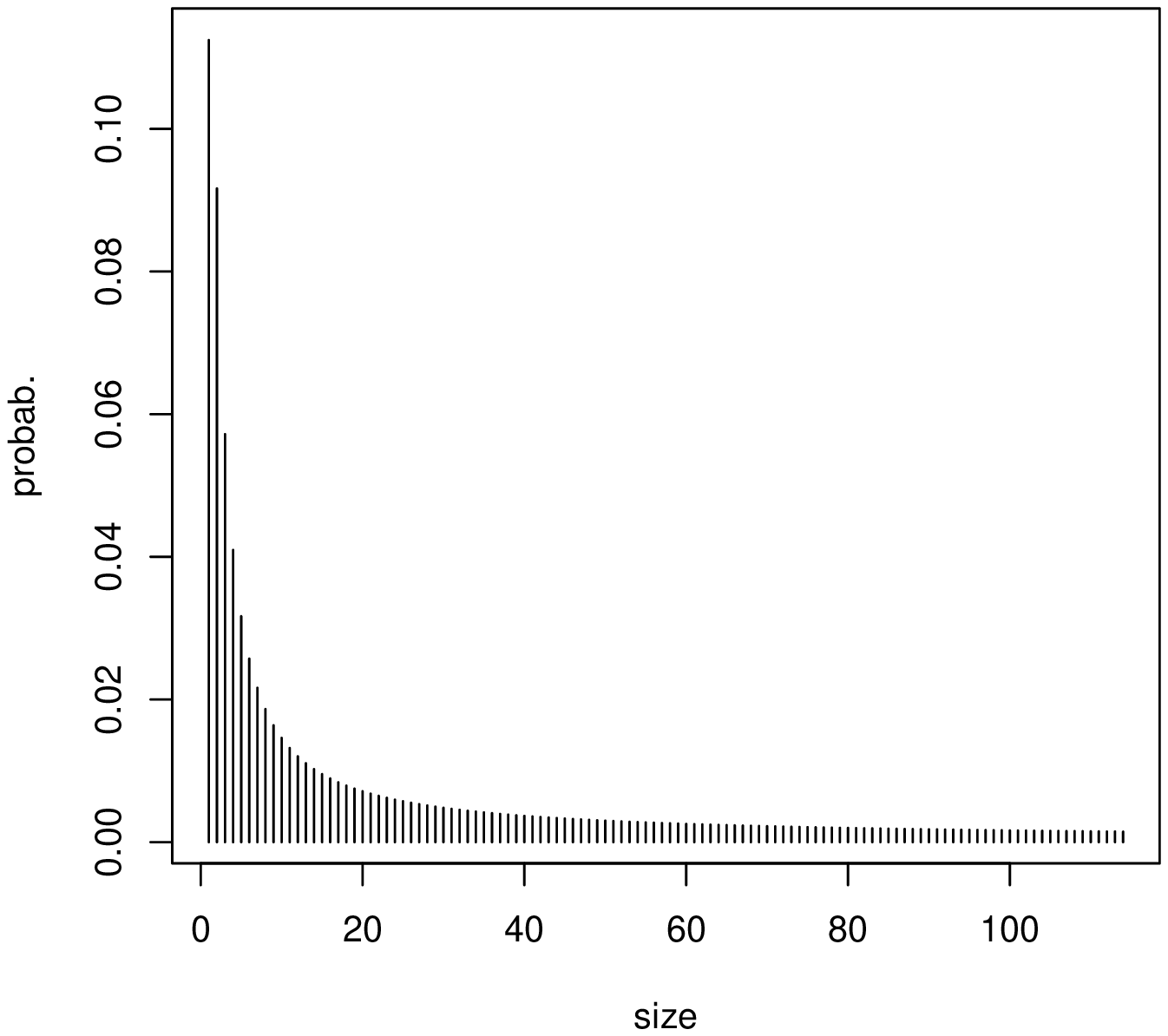}
\includegraphics[width = 6cm]{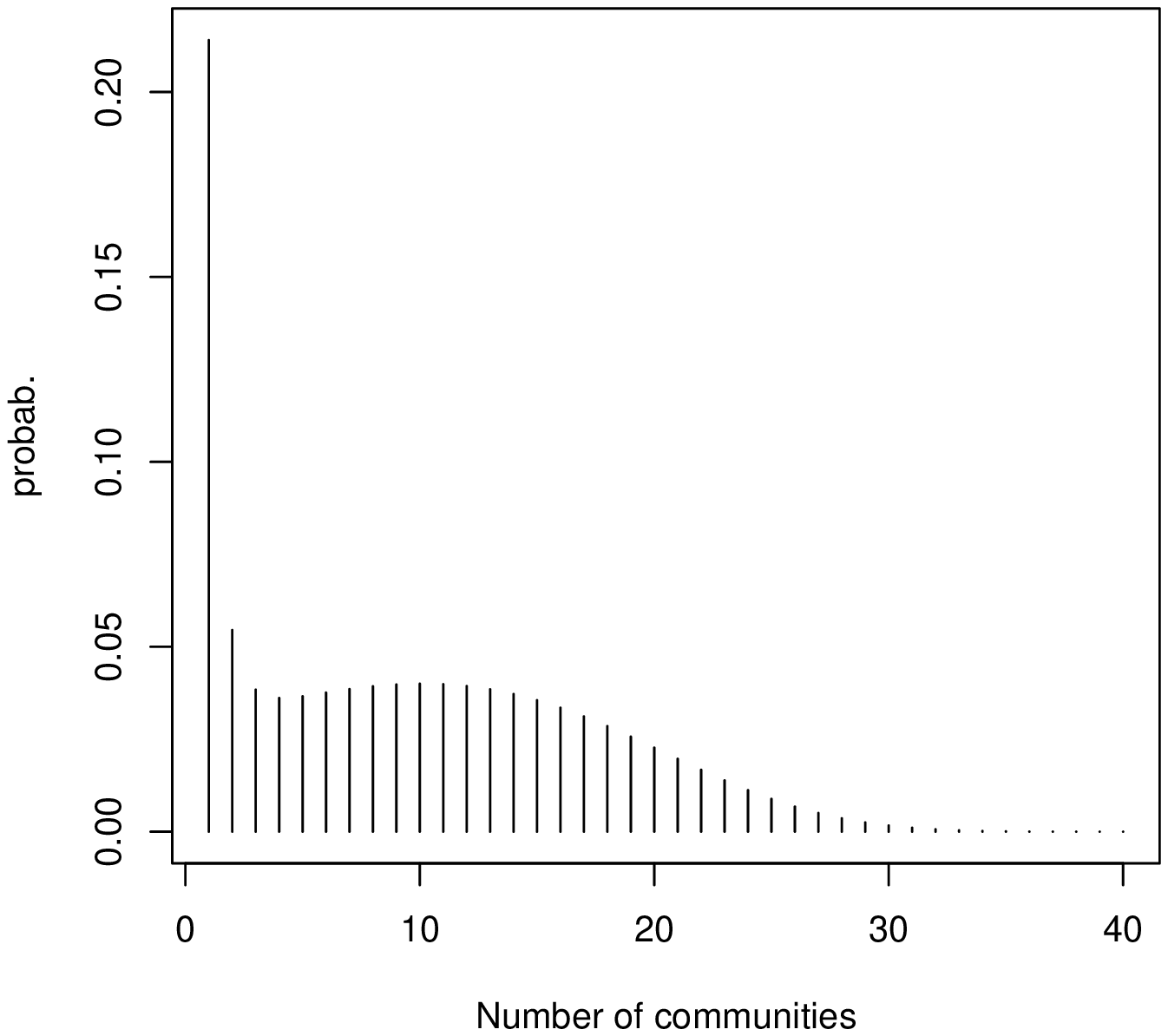}
\caption{Probability distribution of $S_{115}$ (left) and $N_{115}$ (right) 
for $p = 0$ (up) and $p = 0.2$ (down) corresponding to the current American 
football conference network. Data from (Girvan and Newman 2002)}
\end{center}
\end{figure}

\begin{figure}
\begin{center}
\label{fig:wolf}
\includegraphics[width = 6cm]{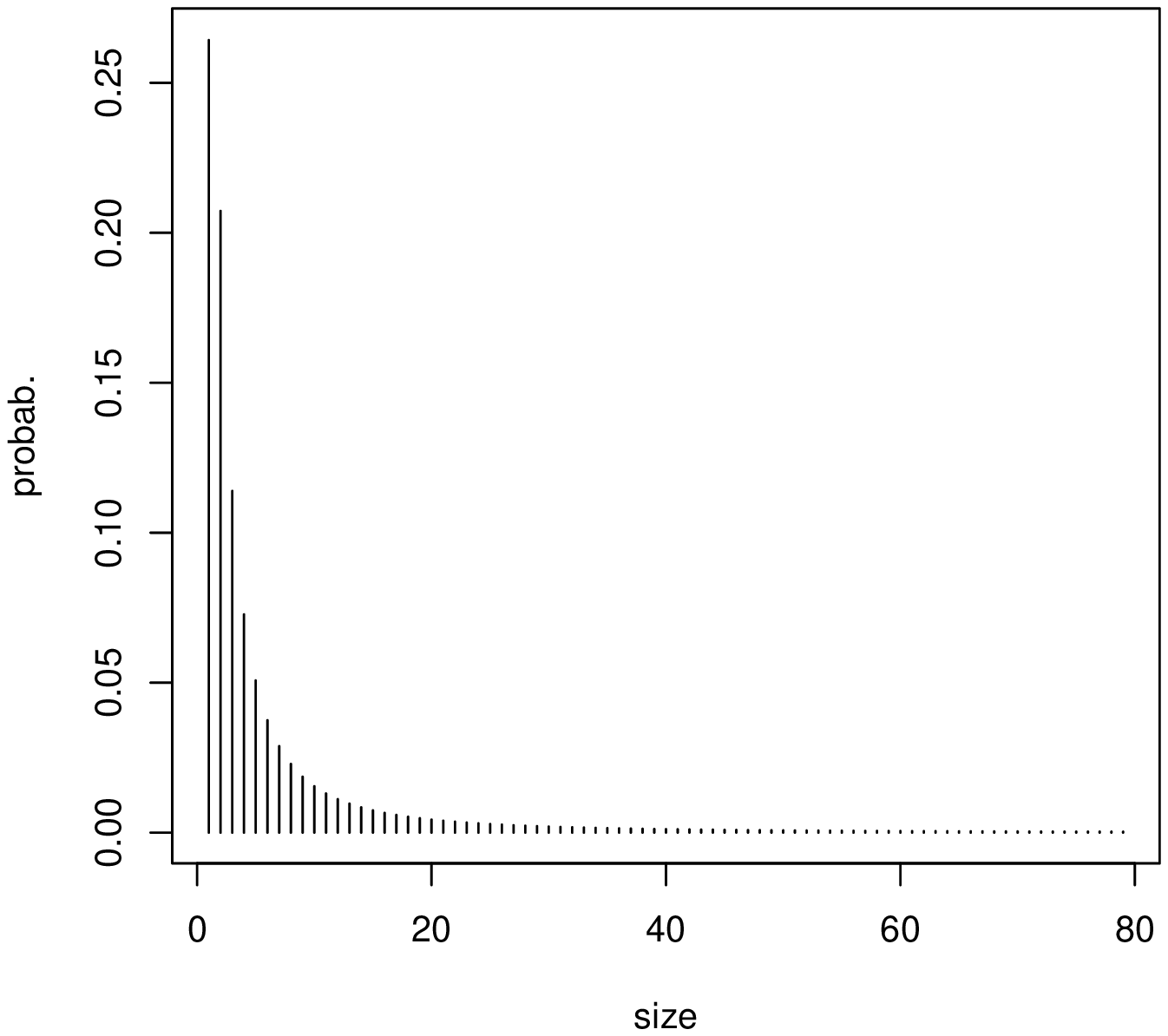}
\includegraphics[width = 6cm]{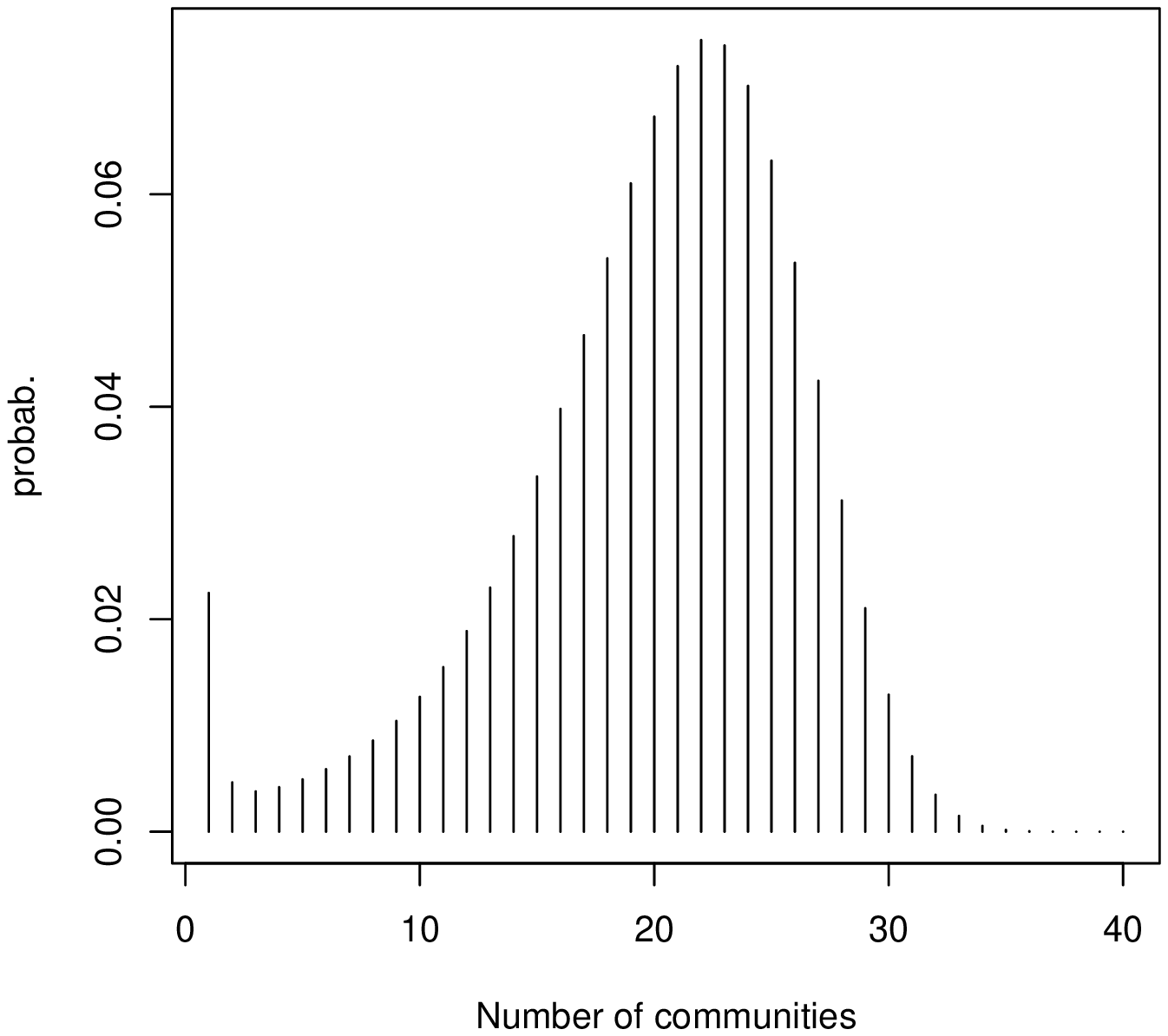}
\includegraphics[width = 6cm]{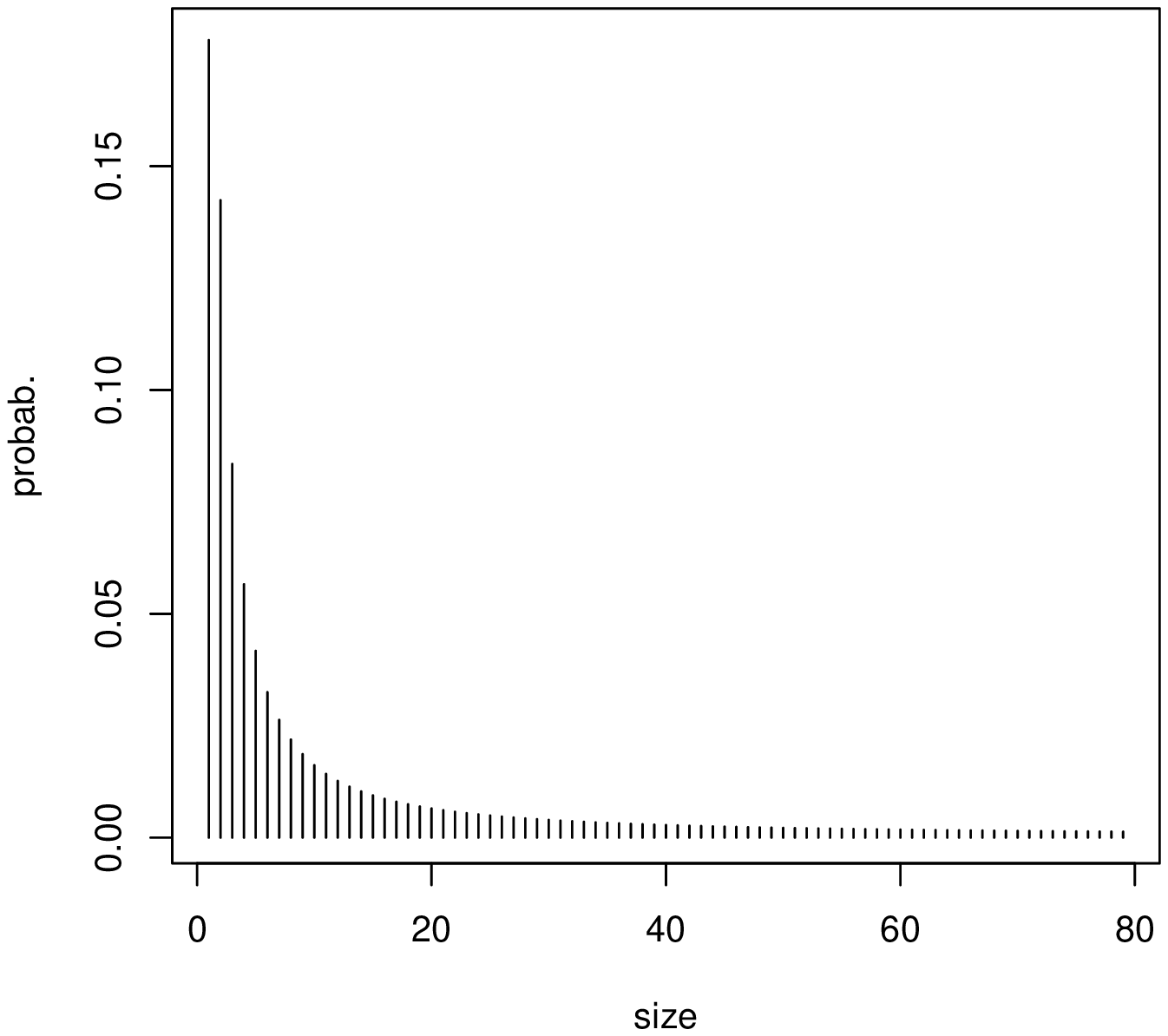}
\includegraphics[width = 6cm]{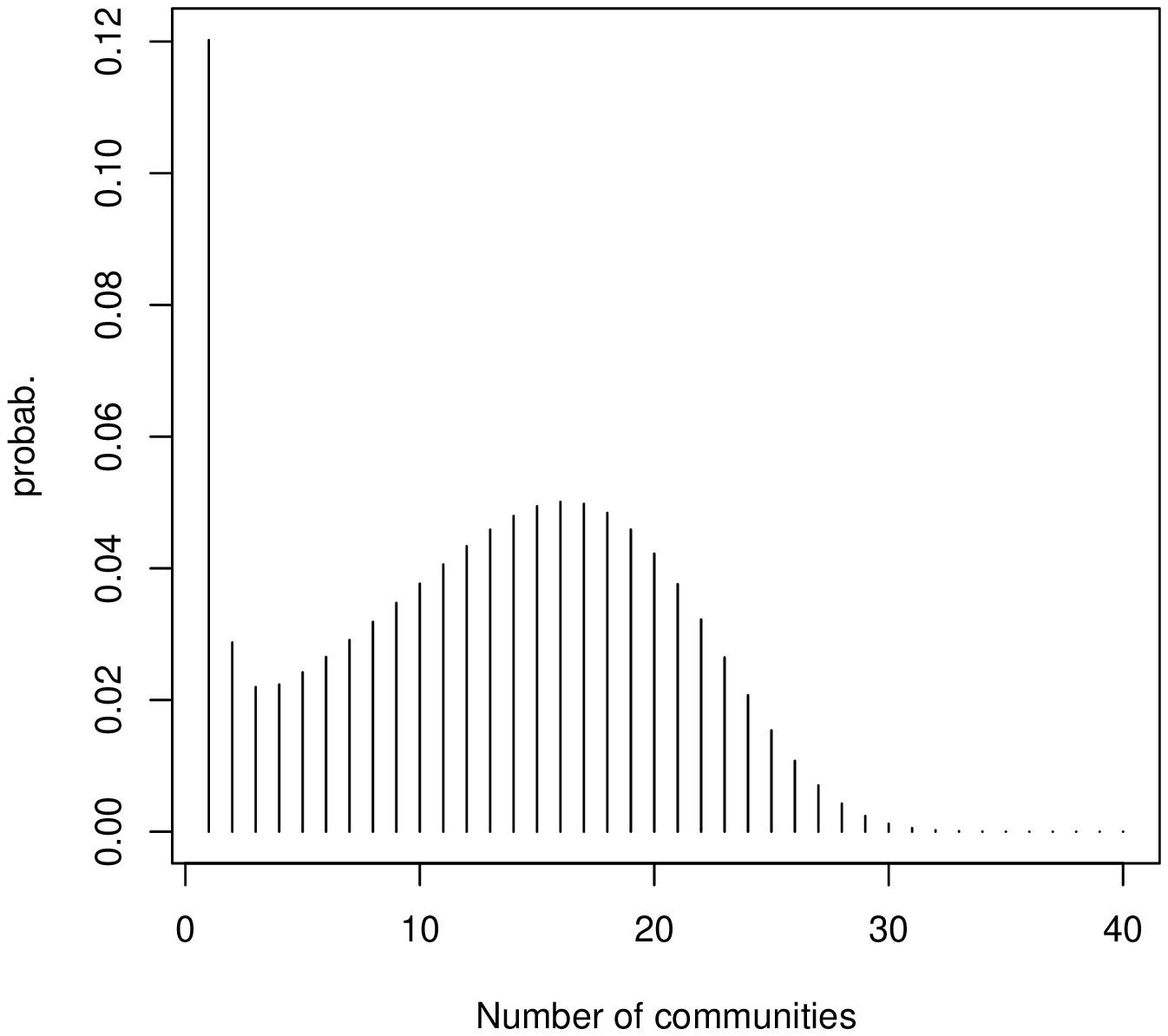}
\caption{Probability distribution of $S_{90}$ (left) and $N_{90}$ (right) 
for $p = 0$ (up) and $p = 0.1$ (down) corresponding to wolf packs data (Smith et al. 2002). }
\end{center}
\end{figure}

The expected number of communities $e_n$ is involved in the following 
recursion
$$
e_{n+1} = \left( 1 - \frac1n \right) e_n + 2\frac{q}{n} e_{n-1} + \frac{p}{n}, \quad n\geq 3
$$
($e_2=e_3 = 1$), that can be solved, and when $n$ grows to infinity, we find that
$$
e_n \sim \left\{
	 \begin{array}{cl}
	    c(p)  n^{1 - 2p} & \textrm{if $p<\frac{1}{2}$} , \\
				& \\
	    \frac{1}{2} \log n & \textrm{if $p=\frac{1}{2}$} , \\
				& \\
	    \frac{p}{2p-1} & \textrm{if $p>\frac{1}{2}$} , \\
	 \end{array} \right.
$$
where $0 \leq p < 1$, 
$$
c(p) \sim \frac{e^{-2q}}{\Gamma(2q)}  \, \left( \frac{1}{q} + q J(p) \right) \, , 
$$
and
$$
J(p) = 2  \int_{0}^{1}  y^2 e^{2qy} (1-y)^{1-2p} dy. 
$$
The exact expression of $c(p)$ is not simple, but we have
$$
c(p) \sim   c + d p + o(p), \quad {\rm as~} p \to 0 
$$ 
where
$$
d = \frac12 {\rm E}_{\rm i} (1,2) + (\gamma - 2)e^{-2} + \log 2 + 1 \approx 0.7747
$$
All the equations given in Sections 3.1-3.2 can be retrieved for $p = 0$. We see that the 
model undergoes a {\it phase transition} at $p = 1/2$ where, for $p<1/2$ the network
has essentially small components, and for $p>1/2$ a giant component may emerge. Examples of 
the above described probability distributions with different network sizes are displayed 
in Fig.\ref{fig:zach}-\ref{fig:wolf}. These graphics are taken from the real 
examples discussed in the next section.



\section{Examples}

\subsection{Collaboration and frienship networks}

\paragraph{Zachary's frienship network} A much-analyzed example of social network is a karate club observed over two years by an anthropologist,
 Wayne Zachary in the 1970s (Zachary, 1977). The network of friendships among the club members has been depicted in a graph by White 
and Harary (2001). The ``karate club" network of Zachary was studied previously by a 
number of other authors in this context (Girvan and Newman, 2002; Zhou, 2003). During the period of 
study, the club splitted in two with those closest to the leader (the karate teacher) 
following him,  and those closest to the administrator as a result of a dispute between two factions.
 Previous studies have found that the fault lines along which the split occurred are readily visible
 in the structure of the network. The network size is $n = 34$ (number of club members). 
Under the mean-field model, we obtained P$(N_{34} \leq 2) = 0.087$, which can be considered
as a one-sided p-value. We computed a one-sided p-value because the model 
is generally more likely to underestimate the true community size than overestimating (here
we have E$[N_{34}] = 7.1 > 2$). The clustering rate was estimated as $\hat p = 0.58$. 
Assuming an error of $p = 1/2$ during the network 
construction, the p-value raised to 0.704. In this and the next examples, the second p-value can be 
interpreted as a type-II error when the perfect clustering model is rejected against an imperfect 
clustering model. In the Zachary's club example, the perfect clustering 
model is rejected at the confidence level $\alpha = 0.087$, but the power of the test is low (around 0.3).

\paragraph{American football conferences}

An interesting example in Girvan and Newman (2002) is the network 
of United States college football, a representation of the schedule of Division I games for the 2000
season: vertices in the graph represent teams. It has a  known community structure, and the reconstructed tree 
published in the original study matches the model presented here quite well (most communities end with an external 
branch). In this example the community structure come from geographical (and historical) relationships between colleges. 
The network has 115 teams and 12 conferences.
Computing the distribution of the number of communities under the mean-field model, we obtained
that P$(N_{115} \leq 16) = 0.063$. The estimated clustering rate was $\hat p = 0.2$, 
 Assuming an error of $p = 0.2$ during the network construction, the p-value increases 
as P$(N_{115} \leq 16) = 0.654$.


\paragraph{Santa-Fe collaboration network.} Girvan and Newman (2002) have also applied their community-finding method to a collaboration 
network of scientists at the Santa Fe Institute, an interdisciplinary research center in Santa Fe, New
Mexico. The 118 vertices in this network represent the largest component of the collaboration graph
among scientists in residence at the Santa Fe Institute during any part of calendar
year 1999 or 2000 and their collaborators. An edge is drawn
between a pair of scientists if they coauthored one or more
articles during the same time period. The algorithm splits the network into six strong communities, 
which lead us to estimate a large clustering rate $\hat p = 0.36$. In this example the neutral model 
is rejected (p-value = 0.029), perhaps due to the fact that the algorithm found surprising groupings, and
the network contains ties between researchers from traditionally
disparate fields. Girvan and Newman conjectured that this feature may be peculiar
to interdisciplinary centers like the Santa Fe Institute.

%
%

\paragraph{A remark about the average community size.} The ecological or socio\-biology literature is not always as formal as we are regarding the average community 
size. The typical community as studied here contains one pre-specified individual. If there are $N_n$ communities
in the sample, then the average 
community size is generally computed as
$$
{\rm Average~community~size} =  \sum_{j = 1}^{N_n} S^j_n / N_n
$$
where the $S^j_n$ are the sizes of the distinct communities within the sample. 
This quantity can be equivalently formulated as $n/N_n$, and its expectation differs from 
$s_n$. Among the examples in the previous section, only the football conferences met the criteria 
of large size and consistency with the mean-field model. In this case, we had $n/N_n = 9.58$
 which was strinkingly close to $s_{115} \approx 2 \log(115) = 9.48$. In general, the bias can be stronger.

\subsection{Groups in social animals}

In the wild, social animals (and especially social carnivores) usually live in 
small sized groups. The groups are given different names according to the 
species. For example lions live in prides, dolphins live in pods, or wolves live 
in packs. A process called {\it kin-selection} was suggested by Hamilton (1964) 
as a mechanism for the evolution of  altruistic behavior, and 
as one of the mechanism that may explain the formation of kin-networks in 
social animal species (Dawkins, 1989; Forster et al., 2006). The process can be sketch as follows. 
Since identical copies of genes may be carried in relatives, a gene that favors altruism 
may become successful provided the reproductive benefit gained by the recipient of the 
'altruistic' act compares favorably to the reproductive cost to the individual performing the 
act. In this comparison, the reproductive benefit gained by the recipient is 
weighted by that the genetical relatedness ($r$) of the recipient, defined as the 
percentage of genes that those two individuals share by common descent. 
Kin-selection involves kin recognition at a basic level, and shares similarities with the 
aggregation model presented in this study. For instance, genetic relatedness
corresponds to a natural measure of closeness between living organisms, and obviously derives 
from a (genealogical) tree. 

In the next paragraphs we compare the mean-field model predictions with published data that 
report precise sample sizes and observed number of groups in Wolves and Lions, 
where kin-selection is often assumed to be acting (see e.g., Rodman, 1981). 
Many workers have proposed the alternative idea 
that the reason social carnivores live in groups, or packs, is because group 
hunting facilitates their acquisition of large prey (Mech 1970; 
Nudds 1978; Pulliam and Caraco 1978). However this idea is not 
shared by all, and recent summaries have argued that communal hunting has 
little power to explain group patterns in felids (Packer et al. 1990) 
and across social carnivores in general (Caro 1994). Beyond this discussion a 
general consensus that kin-selection contributes to the organization and 
evolution of animal social structures remains.

\paragraph{Wolf packs.} Grey wolves ({\em Canis lupus}) are pack-living 
animals with a complex social organization. Packs are primarily family groups.  
Packs include up to 30 individuals, but smaller sizes (8-12) are more common. A 
review of wolf social behavior and ecology can be found in Mech (1970). We use 
data from three sources: The Wolf project of Yellowstone national park which 
annually publishes accurate data on wolf pack sizes (Smith et al., 2002; 2004),
and studies of wolf population recovery after quasi-extinction in Scandinavia
(Wakkaben et al, 2001) and in Alaska (Ballard et al., 1987). When available, the total sample 
size was given as the number of sampled adults (in wolves the number of pups per packs
is usually small). In 2002, $n = 90$ adult wolves were sampled in Yellowstone, 
living in 14 packs. From the mean-field analysis, we obtained that P$(N_{90} \leq 14) = 0.17$.
 The clustering rate was 
$\hat p = 0.11$. Table \ref{tab:1} reports similar results for the year 2004.
In the Alaska, $n = 151$ wolves were sampled, living in 30 packs (number of pups
not known). From the mean-field analysis, we obtained that P$(N_{151} \leq 30) = 0.31$. The clustering rate was 
$\hat p = 0.03$. In Scandinavia, 76 wolves were sampled, living in 12 packs (number of pups
not known). From the mean-field
analysis, we obtained that P$(N_{76} \leq 16) = 0.18$. The clustering rate was 
$\hat p = 0.12$. The last two p-values may be slight underestimates  because the pups were 
included in the sample.


\paragraph{Lion prides.} African lions {\em Panthera leo} live in prides that typically consist of two males, 4-10 females 
and their offspring. The adult females are usually related to one another and 
are group members for life. A review of Serengeti lion behavior and ecology can be found in 
Schaffer (1972). We use recent data from three sources: Selous Game reserve Tanzania (Spong et al., 2002),
Serengeti Tanzania (Packer et al. 2005), and Kafue Park Zambia (Carlson et al, 2004). The study of social 
and genetic structure of Selous Game reserve lions (Spong et al. 2004)
reported the presence of 14 prides, with an average number of 5.6 adults (range 2-9) and 2 males 
in each pride. These observations can be turned into an estimate of 51 females in the sample.
A recent survey of Serengeti lions reports the presence of about one hundred lionesses in the park 
(Packer et al. 2005). 
Based on an average of 6 females per pride, a number of 16-17 prides in Serengeti is consistent 
with the current data. At least 95 adult lions reside in the northern sector of Kafue National Park,
either living in one of 14 prides or roaming as solitary males (Carlson et al., 2004).
Among the adult lions, there are 31 males and 64 females (a sex ratio of 1:2). Nine of the 14 
prides  did not have a sexually mature male residing with them. Pride sizes ranged from 2-14 adult 
animals (mean = 6.4 animals per pride). Of the 17 sexually mature males that were identified, six 
of them associated with prides of females while 11 lived either alone or in all-male
dyads. 

Table \ref{tab:1} reports results for the three samples. As for wolves, the lion samples 
exhibit high p-values and low estimates of the clustering rate $\hat p$. In addition, estimates for Zambia 
may be biased downward because we may have included males. Actual values of female counts would 
exhibit larger p-values, lower clustering rates, and an even stronger agreement with the 
mean-field model.

%

\begin{table}
\label{tab:1}
\begin{center}
\begin{tabular}{ccccc}
\hline
&  & &  &  \\
A. Social networks:  & network & number of        &  p-value & rate   \\
   &  size & communities         &                    & $\hat p$ \\
\hline
&  & &  &  \\
Zachary's & 34 & 2 & 0.087 & 0.58  \\
&  & &  &  \\
Sante-Fe & 118 & 6 & 0.029 & 0.36 \\
&  & &  &  \\
Football  & 115 & 12 & 0.063 &  0.2  \\
&  & &  &  \\
\hline
&  & &  &  \\
B. Social carnivores:  & sample & number of         &   p-value &   rate \\
  &  size & packs/prides       &    & $\hat p$ \\
\hline
&  & &  &  \\
Yellowstone Wolf 2002 & 90 & 14 & 0.17 & 0.11  \\
&  & &  &  \\
Yellowstone Wolf 2004  & 112 & 16 & 0.12 &  0.19  \\
&  & &  &  \\
Alaska Wolf & 151  & 30 &  0.31 & 0.03  \\
&  & &  &  \\
Scandinavian Wolf &  76 & 12 &  0.18 & 0.12  \\
&  & &  &  \\
Zambia Kafue Lions &  95 & 14 &  0.145 &  0.12 \\
&  & &  &  \\
Selous Game Lions &  51 & 13 & 0.64 &  0.00  \\
&  & &  &  \\
Serengeti Lions & 100  & 16 &  0.17  &  0.10 \\
%
\hline
\end{tabular}
\caption{\it Data on community structure. A. Social networks. B. Wolves and lions} 
\end{center}
\end{table}

\section{Discussion}

The mean-field networks presented in this article are rough models of social 
aggregation based on a measure of similarity or kinship. While the network
 clearly reflects an aggregation process, this is also clear that 
the underlying tree model does not account for highly structured interactions. 

The results obtained on several real-world networks and biological data 
have shown that the mean-field model is sufficiently robust to capture 
some essential patterns of group formation, especially if imperfect
clustering is included. In the three examples of social networks, 
the mean-field model was however strongly rejected once (Sante Fe), and weakly rejected twice 
(Zachary's club and Football conferences). Although the test lack power, the fact that 
clustering rates were high suggests that interactions stronger than random may be shaping
these networks. 

The situation was different and perhaps more interesting in social canivore
examples for  which we observed stronger acceptance of the mean-field model.
One perceptible conclusion from 
the results about wolves and lions is that the mean-field model predicts the number of communities 
quite well. This does not contradict the fact that more specific models may better 
explain community structure (e.g., Giraldeau and Caraco, 1993).
Kin-selection is acknowledged to be a major actor of the evolution of social structures 
in wolves and lions, and kin recognition is believed to happen at the same time. Although the mean-field model
 includes kin recognition, 
it actually neglects the effects of selection, or assumes that selection has a very weak impact on 
the shape of the underlying genealogical process. This idea is consistent with mathematical studies of 
selection processes (Neuhauser and Krone 1997, Krone and Neuhauser, 1997). Studying population genetics
models with weak selection, Neuhauser and Krone (1997) actually remarked
that weak selection does not modify the neutral coalescent topology significantly.

Although wolves/lions data agree with the perfect clustering mean-field model predictions, 
there are other social species for which the fit may be poorer. This may be the case of 
fish schools or large ungulate herds, where other models of group formation may be more
appropriate (e.g., Bonabeau et al, 1999). For example, in an aerial survey of known and 
suspected wild camels habitat {\em Camelus bactrianus},  Reading et al. (1999) estimated 
group density and population size of large ungulates in the south-western Gobi Desert in Mongolia. 
They observed 277 Wild camels in 27 groups, which leads to a strong reject of 
the mean-field model (p-value = 0.026). The same is also true for Buffalos that live in 
herds much larger than the community size predicted by the mean-field model.

In summary, we have presented a  mean-field analysis of 
community structure in tree-derived networks that includes 
an attachment process to closest vertex deduced
from the tree. Our model is reasonably simple, and we have obtained exact results 
about the typical community sizes and the numbers of communities. 
While community structure in studied social networks has exhibited weak departures 
from the perfect clustering mean-field model, predictions of imperfect clustering 
models with higher clustering rates are consistent with the data. This suggests that  
stronger interactions than random may be present in these networks. 
Examples of social animals have provided a better fit to the mean-field model. 
In populations evolving altruistic behavior, the results suggest that kin-recognition 
may contribute to shape community  structure more significantly than natural selection 
does itself.

\bibliographystyle{empty}

\section*{Mathematical Appendix}   

\subsection*{Results for perfect clustering}

\paragraph{Asymptotics of $p(x)$ for large $x$ (Distribution of $S_n$).} We have 
$$
e^{-1}=\sum_{k=2}^{\infty}(-1)^k/k!
$$
and then 
$$
e^{-1}-e(x)=\sum_{k=x+1}^{\infty}(-1)^k/k!
$$
We deduce that 
$$
e^{-1}-e(x)=(-1)^{x+1}/(x+1)! +o(1/(x+1)!),
$$
and plugging this into the formula for $p(x)$, we obtain
$$
p(x) ~ \sim  \frac{2}{(x-1)(x+1)} , \quad {\rm as~} x \to \infty
$$

\bigskip

\paragraph{Second moment of $S_n$.} Regarding the second order moment 
$s_n^2 = E[S_n^2]$, we find that the difference 
$u_{n + 1} = s^2_{n+1} - s^2_n$
satisfies 
$$
u_{n+1} = - \frac1n (u_n - 2(2n +1))
$$
which can also be solved, and yields  
$
u_n \sim  (4 A_{n+1} - 2 A_n) / n n! \, \sim 4 
$
In conclusion we have 
$$
s_n^2 \sim 4 n 
$$

\bigskip

\paragraph{Higher moments of $S_n$.}
Let us denote by $\phi_{n}(t)=E[e^{tS_n}]$ the moment generating function of 
$S_n$. Then using the formula of conditional probabilities we obtain a 
functional recurrence equation for $\phi_n(t)$ 
\begin{equation}
\label{eq:phi} \phi_{n+1}(t) = (1-\frac{1}{n})\phi_{n}(t) + \phi_{n-
1}(t)+\frac{2}{n}e^{tn}(e^{t}-1).
\end{equation}
Now we set $f_{n}(t) = 2 e^{tn}(e^{t}-1)/n$, and see that the derivatives of 
$f_n(t)$ at $t=0$ are equal to 
$$
f_{n}^{(k)}(0)=\frac{2(n+1)^{k-
1}}{n}+2((n+1)^{k-1}-n^{k-1})$$
for all $k \geq 1$. The $k$th moment of $S_{n}$, noted $s_{n}^{(k)}$ is equal to 
$\phi_{n}^{(k)}(0)$. Thus it solves the equation $s_{n+1}^{(k)}=( 1 - s_{n}^{(k)}/n )+  s_{n-
1}^{(k)}/n + f_{n}^{(k)}(0)$. If we denote $u_{n}^{(k)} = s_{n+1}^{(k)}-
s_{n}^{k}$, then for all $n \geq 2$, $k \geq 1$, we have $u_{n}^{(k)}=- u_{n-
1}^{(k)}/n + f_{n}^{(k)}(0)$. Using the Newton's binome formula, we check that 
$f_{n}^{(k)}(0) \sim 2k n^{k-2}$ for large $n$. So we have $u_{n}^{(k)} + u_{n-
1}^{(k)}/n=f_{n}^{(k)}(0) \sim 2k n^{k-2}$. Thus 
$$u_{n}^{(k)} \sim 2kn^{k-2}$$
Finally, when $n$ grows to infinity and $k \geq 2$, we have
$$
s_{n}^{(k)} \sim  \frac{2k}{k-1}n^{k-1} 
$$
as $n$ grows to infinity.

\paragraph{Mean number of communities.} The sequence $f_n$ 
defined by $f_n=e_{n+2}$ satisfies
\begin{equation}
\label{f} (n+1)f_n=(n-1)f_{n-1}+2f_{n-2}
\end{equation}
where the recursion is initialized as $f_0=f_1=1$. Since the
equation satisfied by $f_n$ is a linear recurrence equation of order  2, $f_n$ 
is a linear combination of two independent solutions of eq. (\ref{f}). One 
straightforward solution of eq. (\ref{f}) is
$$
u_n=n+4 , \quad n \geq 2
$$
with $u_0=4$ and $u_1=5$.

The rest of the proof is devoted to finding a solution $v_n$ of eq. (\ref{f}) 
with initial values $v_0=0$ and $v_1=1$. Let us denote $h$ the generating 
function of $v_n$. Then we have
$$
h(x)=\sum_{n=2}^{\infty} v_n x^n.
$$
Using equation (\ref{f}), we find that $h$ is a solution of the following
differential equation
$$
h(x)(1-2x^2-x)=h'(x)(x^2-x)+2x^2+2x^3
$$
Solving the above differential equation leads to
$$
h(x)=\frac{e^{-2x}-1}{(x-1)^2 x} + \frac{-x^4+2x^3-2x^2+2x}{(x-1)^2
x}
$$
Using the Taylor expansion of $h$, we find that
$$ v_n=(n+2)-\sum_{k=0}^n \frac{(-2)^k (n+1-k)}{(k+1)!}\;\;, n\geq3
$$
Since $f_n$ is a linear combination of $u_n$ and $v_n$
$$
f_n=\frac{1}{4}u_n - \frac{1}{4}v_n,
$$
we find that the expected value of the number of
communities is given by
$$
e_n=\frac{1}{2} +\frac{1}{2} \sum_{k=0}^{n-2}
\frac{(-2)^k(n-1-k)}{(k+1)!} \, , \quad n \geq 4.
$$
This can be rewritten as
$$
e_n=\frac{1}{2} +\frac{1}{2} \left((n-1)\sum_{k=0}^{n-2}
\frac{(-2)^k}{(k+1)!}-\sum_{k=0}^{n-2} \frac{(-2)^k k}{(k+1)!}
\right).
$$
Since the rest (beginning with term $n$) of convergent alternating
series is dominated by the absolute value of the $(n+1)^{th}$ of the
series, we have
$$
e_n=\frac{1}{2}-\frac{1}{4} \left( (n-1) \sum_{k=0}^{\infty}
\frac{(-2)^k}{(k+1)!}-\sum_{k=0}^{\infty} \frac{(-2)^k k}{(k+1)!}
+O(2^n/(n-1)!) \right).
$$
For large $n$, this leads to
$$
e_n = \frac{1-e^{-2}}{4}(n-1) + \frac{3(1-e^{-2})}{4} + O(2^n/(n-1)!).
$$

\bigskip

\subsection*{Results for imperfect clustering}

Here the results are presented in a reversed order compared to the text. We first prove results 
for the mean number of communities (easier), and give a sketch of proof for the mean 
community size.

\paragraph{Mean number of communities.}
Assuming imperfect clustering at rate $p$, the expected number of communities 
solves the following equation:
\begin{equation}
\label{eq:en} (n+1)e_{n+2} = ne_{n+1} + 2 q e_{n} + p
\end{equation}
with the initial values $e_{2} = e_{3} = 1$. For $p \neq \frac{1}{2}$ we set $f_{n} = e_{n+2} - p/(2p-1)$,
and obtain that
\begin{equation}
\label{eq:fn} (n+1)f_{n} = nf_{n-1} + 2 q f_{n} 
\end{equation}\newline
Let us denote by $h_{f}(x)$ the generating function of $f_{n}$
$$
h_{f}(x) = \sum_{n=2}^{\infty}f_{n}x^{n}.
$$
We first investigate a solution $f_{n}$ to equation (\ref{eq:fn}) 
with $f_{0} = 0$ and $f_{1} = 1$.
The generating function solves the following differential equation
\begin{equation}
\label{eq:y1} (1-x - 2q x^{2})y  = (x^2-x) y' + 2x^2 +  2q x^3
\end{equation}\\
The analytical solution of equation (\ref{eq:y1}) is given by 
$$
h(x) = \frac{(x-1)^{-2q} e^{-2qx}}{x}\frac{e^{-2i\pi p}}{q}+\frac{-qx^2-1}{qx}
$$
A series expansion leads to  
$$
f_{n} =\frac{1}{q}\frac{\Gamma (n+1+2q)}{\Gamma(2q)(n+1)!}+\frac{1}{q}\sum_{i=0}^{n}\frac{\Gamma(n-i+2q)}{\Gamma(2q)(n-i)!}\frac{(2q)^{i+1}}{(i+1)!}
$$
for all $n\geq 2$. Using the fact that $\frac{\Gamma(n+2q)}{n!} \sim n^{1-2p}$, we find that
$$
f_{n} \sim \frac{e^{-2q}}{q\Gamma(2q)}n^{1-2p}
$$
Now we seek a solution to equation (\ref{eq:fn}) with $f_{0} = 1$ and $f_{1} = 0$.
The generating function can be involved in the following differential equation 
\begin{equation}
\label{eq:y2} (1-x-2q x^{2})y = (x^2-x)y' + 2 q x^2
\end{equation}
The analytical solution of equation (\ref{eq:y2}) is 
$$
h(x)=\frac{1}{x}\int_{0}^{x}2(y-1)^{1-2p}y^2(p-1)e^{2yq}dy \frac{e^{-2xq}}{(x-1)^{2q}}
$$
Using Darboux' result, Theorem 4.12 in (Sedgewick and Flageolet, 1996), we find that 
$$
f_{n} \sim e^{-2q} n^{1-2p}\int_{0}^{1}2(1-y)^{1-2p}y^2qe^{-2yq}dy 
$$
Finally, adding the two previous solutions, we find that the solution of the original problem
is equivalent to
$$
e_{n} \sim \left\{
	   \begin{array}{ll}
	      \frac{e^{-2q}}{\Gamma(2q)} \left( \frac{1}{q} + q I(p) \right) \, n^{1-2p} & \textrm{if $p<1/2$}\\
	      \frac{p}{2p-1} & \textrm{if $p>1/2$}
	   \end{array} \right.
$$
where we have set 
$$
I(p) = 2 \int_{0}^{1} (1-y)^{1-2p} y^2 e^{2qy}dy. 
$$
In the case $p=1/2$, $e_{n}$ satisfies the equation 
$$
(n+1)e_{n+2} = ne_{n+1} + e_{n} + \frac{1}{2}
$$
Denoting $u_{n} = e_{n+1} - e{n}$, we obtain that $u_{n} \sim 1/2n$ and 
thus we have $e_{n} \sim \log n / 2$.

\paragraph{Community size.} Assuming imperfect clustering at rate $p$, the mean community size 
$s_{n}$ satisfies the following recursive equation
$$
s_{n+1} = (1-\frac{1}{n})s_{n} + \frac{q}{n}s_{n-1} + 2p + \frac{2q}{n} 
$$
Let us denote 
$$ u_{n} = \frac{2p}{1+p}n, $$ 
and 
$$
f_{n} = s_{n} - u_{n}.
$$
We remark that $u_{n}$ satisfies the following equation
$$
u_{n+1} = (1-\frac{1}{n})u_{n} + \frac{q}{n}u_{n-1} + 2p + \frac{2pq}{n(1+p)}.
$$
Thus, we have
$$
n f_{n+1} = (n-1)f_{n} + q f_{n-1} + (2q - \frac{2pq}{1+p})
$$
Applying the same method as in the previous paragraph for $p>0$, we obtain that 
$$
f_{n} = s_{n} - u_{n} \sim K
$$
where $K$ is a constant term. Then this is routine to conclude that  
$$
s_{n} \sim u_{n} = \frac{2p}{1+p}n ,
$$
which is also in good agreement with numerical values for moderate $n$.
\end{document}